\documentstyle[12pt,titlepage]{article}

\def\bib{\bibitem}
\def\be{\begin{equation}}
\def\ee{\end{equation}}
\def\beqar{\begin{eqnarray}}
\def\eeqar{\end{eqnarray}}
\def\barr{\begin{array}}
\def\earr{\end{array}}

\def\gsim{\:\raisebox{-0.5ex}{$\stackrel{\textstyle>}{\sim}$}\:}
\def\and{\qquad {\rm and } \qquad}
\def\ie{ {\it i.e.} }
\def\eg{ {\it e.g.} }
\def\arrowA{\displaystyle {\mathop{\longrightarrow}_{A^*(q)}}}
\def\arrowH{\displaystyle {\mathop{\longrightarrow}_{H^*(q)}}}
\def\arrowz{\displaystyle {\mathop{\longrightarrow}_{Z^*}}}
\def\etsl{$\not${\hbox{\kern -2.0pt $E_T$}} }

\begin{document}
\thispagestyle{empty}
\begin{flushright}
TIFR--TH/92--55 \\
November 1992
\end{flushright}

\vspace{5ex}
\begin{center}

{\large \bf Higgs-mediated heavy neutrino}\\
{\large \bf  pair-production at $pp$ supercolliders}

\bigskip
\bigskip
{\sc
   Debajyoti Choudhury$^{1,}$\footnote{debchou@tifrvax.bitnet},
   Rohini M. Godbole$^{2,}$\footnote{rohini@tifrvax.bitnet}
   and
   Probir Roy$^{1,}$\footnote{probir@tifrvax.bitnet}
   }

\bigskip
$^{1.}$ {\it Tata Institute of Fundamental Research, Homi Bhabha Road,
Bombay 400 005, India.} \\
$^{2.}$ {\it Physics Department, University of Bombay, Vidyanagari, Bombay 400
098, India.}

\vspace{5ex}
{\bf Abstract}\\

\medskip
\begin{quotation}
Gluon fusion into a very heavy neutrino pair by Higgs exchange
is shown to lead to substantial production cross sections at $pp$
supercolliders even without any extra generation of quarks.  Rates are
calculated for scalar as well as pseudoscalar Higgs.  The angular
correlation between dileptons emerging from the decays of the neutrinos
shows distinctive features for Dirac and Majorana neutrinos as well as for
scalar and pseudoscalar Higgs.
\end{quotation}
\end{center}
\newpage

The production and subsequent decays of very heavy neutrinos
(generically denoted $N$ with $M_N > M_Z/2$) in forthcoming
colliders has attracted a certain amount of recent interest
[1-6].  Such particles are predicted by many theories --- \eg a
4th generation extension \cite{AD} of the Standard Model (SM),
left-right symmetric theories \cite{LR} as well as $SO(10)$ or
$E_6$-based grand unified models \cite{GUT}.  However, their
properties are largely unconstrained by low energy phenomenology
or astrophysical/cosmological considerations  so long as they
are unstable, stable ones being cosmologically disfavoured
\cite{astro}.  Thus
the best hope of detecting them rests on their copious
production at high energy supercolliders and on the observation
of their decays into distinctive final states.

The requirement of a reasonable signal to background ratio in
the above process points towards reactions induced by neutral
currents rather than those effected by charged currents
\cite{buch}.  Taking this cue, we concentrate on the gluon
fusion mechanism in a $pp$ supercollider producing a $Z$ boson
or a spin zero Higgs particle (off-shell, in general,) which
later goes into a pair of very heavy neutrinos that decay within
the detector.  The naive expectation would be for the
quark-antiquark annihilation mechanism of Drell and Yan to
produce the dominant contribution.  Yet, this is belied
\cite{LR} on account of two reasons.  (1) The density of gluons
in the proton is far higher than that of quarks or antiquarks in
the kinematic region of interest.  (2) The quark mass appears as
a factor in the coupling of quarks to the Higgs as well as to
the longitudinal component of the $Z$ suppressing the light
quark contributions.  Thus gluon fusion (through a heavy quark
triangle) into a $Z$ or Higgs and then into an $N \overline N$
pair is the dominant production process at high energies.  Ref
\cite{dicus} has already discussed the $Z$-mediated process.
Our focus in this communication is on the Higgs contribution to
the gluon-fusion mechanism and some of its interesting
properties.

We assume that the heavy neutrinos decay within the detector
into a light charged lepton ($e,\mu,\tau$) by charged current
interactions, \ie  $N(\overline N) \rightarrow \ell (\overline
\ell)$ + a real or virtual $W$ converting into a
lepton--neutrino pair or into jet(s). Universality
constraints on such couplings are rather weak at present
\cite{universal}.  For heavy neutrinos belonging to a fourth
generation, a generalized GIM mechanism would operate and these
would be the only decay modes.  For the case when $N_L$ is an
$SU(2) \times U(1)$ singlet (and/or $N_R$ is part of an $SU(2)$
doublet) there could also be a \cite{schechter} neutral
current induced decays: $N(\overline N) \rightarrow
\nu_L(\overline \nu_L)$ + a real or virtual $Z$.  However, each
such decay would generate a very large
\etsl associated
with a possibly reconstructible $Z$.  Thus the corresponding
events would have rather spectacular signatures.  Our concern
here is with the less spectacular but nevertheless interesting
signals generated from the charged current induced decays of $N,
{\overline N}$.

Thus, when $N$ is Majorana, we would have a
like-sign dilepton pair without
\etsl and with jets,
arising from the semi-hadronic decays of the pair.  For a Dirac
pair, though, one has to be more careful because of the
background from the production and semi-leptonic decays of a
$t\overline t$ pair.  The best signal to trigger on now is the
one consisting of the leptonic decays of both the heavy
neutrinos.  Despite the loss in terms of the branching fraction,
one gains enormously in the cleanliness and uniqueness of the
signal.  The final state is characterized by four charged
leptons,
\etsl and no jets.  The choice of this {\it
hadronically quiet} ~mode then, at a single stroke, gets rid of
the potentially troublesome background from a $t \bar t$
production. With this constraint and the further realization
that the 4--lepton signal from $t \bar t$ is suppressed
by the branching fractions and that two of these leptons have. on the
average, much lower $E_T$, one sees easily that such backgrounds
need not be worried about. The prominent background then arises
from the production and decays of double $Z$'s. With a cut on
\etsl
and additionally a check on the unlike sign dilepton invariant
mass distributions, this can be tackled.

We assume a phenomenologically interesting mass-range of 50 --
500 GeV for a generic $N$.  The mass-range of the exchanged
Higgs is also taken to be similar.  For simplicity, consider
either a pure scalar or a pure pseudoscalar coupling of the
latter to the heavy neutrino pair.  Such is the case in several
interesting models --- in particular --- in the minimally
extended SM with two Higgs doublets (with or without
supersymmetry).  Left-right symmetric models, also have this
property in consequence of the required absence of any tree
level flavour-changing neutral current.  We need not, however,
confine ourselves to any particular model at this stage and will
come to specific model-dependence later.

The above assumptions restrict the relevant Yukawa Lagrangian to
\be
 {\cal L}_Y = \left( G_F/\sqrt {2} \right)^{1/2}
    \left[\sum_i m_i
      \left\{ (h_S)_i \overline Q_i Q_i H
             + (h_p)_i\overline Q_i \gamma_5 Q_i A
        \right\}
      + m_N \left\{ h'_S \overline N N H
                   + h^\prime_p\overline N\gamma_5 N A
             \right\}
    \right].
                      \label{lagrngian}
\ee
The index $i$ here runs over the heavy quarks $Q = (t,b)$.
Terms involving the lighter quarks and ordinary leptons have
been suppressed.  $H$ and $A$ are scalar and pseudoscalar fields
respectively while $h_{S,P} (h'_{S,P})$ are real constants, their
departure from unity reflecting any possible deviation from the
SM couplings caused by either fermion mixing or Higgs mixing.

In many models (\eg, one with two Higgs doublets), there are
more than one neutral Higgs. Of these some have scalar couplings
with the fermions, and we shall generically call them $H$. There
are also those with pseudoscalar couplings and are generically
designated $A$. Their vertices with two gluons, at one--loop
order, is generated through heavy quark triangle diagrams.
The gluon-gluon-scalar \cite{LR} vertex (with
on-shell gluons of four momenta $k_1$ and $k_2$, $k_1 + k_2 =
q$, but generally off-shell Higgs) is :
\be
{\cal S}^{ab}_{\mu\nu} (k_1,k_2,q = k_1+k_2) =
   \displaystyle
    \frac{\alpha_S(q^2)}{\pi q^2}~\delta^{ab}
    \left( k_{1\nu}k_{2\mu} - \frac{q^2}{2} g_{\mu \nu} \right)
    \sum_i (h_S)_i m_i
      \left\{ \left(1 - \frac{4}{a_i} \right)~f(a_i) -
             2\right\}.
                         \label{ggH}
\ee
In (\ref{ggH}) $a_i = q^2/m^2_i$ and$\displaystyle ^1$ $f(a) =
\int^1_0~dx~x^{-1}
\ln [1 - x(1-x)~a+i\epsilon]$.  Furthermore, $\alpha_S (q^2)$ is the
running QCD fine structure constant.  Similarly, for the
gluon-gluon-pseudoscalar vertex, we have
\be
  {\cal P}^{ab}_{\mu\nu} (k_1,k_2;q = k_1 + k_2)
     = i~ \frac{\alpha_S(q^2)}{\pi q^2}~\delta^{ab}
        ~k_1^\lambda k_2^\sigma \epsilon_{\lambda\mu\nu\sigma}
          ~\sum_i (h_P)_i~m_if(a_i).
                         \label{ggA}
\ee

A comparison of the above expression with that for the $ggZ^*$
vertex \cite{dicus,GUT} is instructive.  The latter has an
enhancement \cite{GUT} at high energies because of Yang's
theorem, yet it is unable to dominate over the former.  The key
reason for this fact lies in the group theoretic nature of the
$Q_i\overline{ Q_i} Z$ coupling.  The leading term in the $ggZ^*$
vertex comes proportional to $Tr(T_{3L})$ and hence vanishes as
a part of the anomaly cancellation.  The next-to-leading term is
numerically much smaller.  Such is definitely not the case for
either of the $ggH$ and $ggA$ vertices.  Between the two,
though, the pseudoscalar vertex is less affected by such
cancellations in comparison with the scalar one.  Thus, for
equal coupling strengths, the pseudoscalar amplitude is somewhat
larger.

The differential cross section for the process $g(k_1)~g(k_2)
\arrowH N(p_1,S_1)~{\overline N}(p_2,S_2)$, with $ S^\mu$ standing
for the spin pseudovector, is
\be
\barr{rl}
\displaystyle \frac{ d\hat\sigma^D(H)}{d\Omega_N} =
   & \displaystyle
    \frac{ G_F^2}{16 \pi^2q^2}~ \left(\frac{\alpha_S}{\pi}\right)^2
      |h'_S|^2~
      \left|\sum_i (h_S)_i m^2_i
         \left\{ 1 + \left( \frac{2}{q_i} - \frac{1}{2} \right)~
           f(a_i) \right\} \right|^2
      M^2_N \left(1 - \frac{4M^2_N}{q^2} \right)^{1/2}
         \\
   & \displaystyle
    \left\{ (q^2 - m^2_H)^2 + \Gamma^2_H m^2_H \right\}^{-1}
    \left[ (1 - S_1 \cdot S_2)~(p_1\cdot p_2 - M^2_N)
          + p_1\cdot S_2~ p_2\cdot S_1
    \right].
\earr
                      \label{H-diff.c.s.}
\ee
In contrast, for $g(k_1)~g(k_2) \arrowA N (p_1,S_1)~\overline
N(p_2,S_2)$, it is
\be
\barr{rl}
\displaystyle \frac{ d\hat\sigma^D(A)}{d\Omega_N} =
   & \displaystyle
    \frac{ G_F^2}{64 \pi^2q^2}~ \left(\frac{\alpha_S}{\pi}\right)^2
      |h'_P|^2~
      \left|\sum_i (h_S)_i m^2_i  f(a_i) \right|^2
      M^2_N \left(1 - \frac{4M^2_N}{q^2} \right)^{1/2}
         \\
   & \displaystyle
    \left\{ (q^2 - m^2_A)^2 + \Gamma^2_A m^2_A \right\}^{-1}
    \left[ (1 + S_1 \cdot S_2)~(p_1\cdot p_2 + M^2_N)
          - p_1\cdot S_2 ~ p_2\cdot S_1
    \right].
\earr
                      \label{A-diff.c.s.}
\ee
In (\ref{H-diff.c.s.}) and (\ref{A-diff.c.s.}) $\hat\sigma^D$
denotes the subprocess cross section for producing a Dirac pair
$N,\overline N$ out of gluon-gluon fusion.  For a Majorana pair
with similar couplings, one would instead have \cite{AD}
$\hat\sigma^M = 2\hat\sigma^D$.  The observable cross-section
$\sigma(pp \rightarrow N\overline N + \ldots)$ in the laboratory
has to be obtained by folding the above expressions with the
gluon distribution functions for which we use the $GGR$
densities, as parametrized in Ref.\cite{ggr} using the input of
DFLM \cite{dflm}.

The spin-summed integrated cross sections for the laboratory
process $pp
\rightarrow N\overline N + X$ have been plotted in Figs.
($1~a,b$), for
various values of $m_H$.  We have taken two representative
cases.  In Case I we have just three quark generations and have
included only the $t-$ and $b-$ contributions with $m_t$ chosen
to be $160 \:GeV$ (a change of $20\: GeV$ in $m_t$ would affect
the cross--section at the level of only a few percent, an
uncertainty inherent in the parametrizations for parton
structure functions).  In contrast, Case II includes an additional
heavier fourth generation with nearly degnerate quark masses
$m_U \sim m_D \sim 400$ GeV.  In order to be definite, we
concentrate here on very heavy Dirac neutrinos and take the
mixing factor at the $HN\overline N$ vertex to be unity. [This
is the case not only for a fourth generation $N$, but
also occurs for those $E_6$ GUTs where the same Higgs
couples to both the $SU(2)_L \times U(1)_Y$ singlet neutrinos
and the extra $Q = -1/3$ quarks present in the fermion 27-plet.]
The corresponding plots for the pseudoscalar case occur in Figs.
($2~a,b$). The result could be viewed from  a different
perspective if the cross--sections were plotted as  functions of
$m_H$ instead (Fig. 3). This provides us with a ready estimate
of the range of $m_H$ for which a heavy neutrino of a given mass
and coupling would be observable at the SSC.

We want to emphasize the point that the {\it cross sections are
quite large even without the introduction of 4th generation
quarks}.  In fact, for a very wide region of the parameter
space, these are orders of magnitude larger than that for the
$Z$-mediated process.  This is true even when one is
significantly away from the peaks (corresponding to the
$s$-channel Higgs resonance) in Fig. 1-3.  This result, for
Majorana neutrinos (since only those were considered there), was
basically contained in the work of Datta and
Pilaftsis \cite{AD}.  However, in all realistic seesaw models of
Majorana mass-generation, there is a small mixing angle in the
$HNN$ coupling.  This led to an effective suppression of the
cross section in their case.  Thus, both in Refs.
\cite{dicus,AD}, an extra generation of heavy quarks had to be
brought in to generate measurable rates.  Such a suppression
factor need not be present for a heavy Dirac neutrino.

An additional feature of our investigation is the pseudoscalar
channel.  Given the not unnatural assumption of comparable heavy
neutrino coupling strengths for the scalar and the pseudoscalar
Higgs, it is evident from Fig. 1-2 that the cross section for
the pseudoscalar exceeds that for the scalar by almost an order
of magnitude.  This is a new result of considerable significance
--- especially in the context of two Higgs doublet models (with
or without supersymmetry).  Of course, this observation gets
vitiated if the Higgs has simultaneous scalar and pseudoscalar
couplings; but the latter is incompatible with the principle of
natural flavour-conservation \cite{lee-wein} and we ignore it.

The next issue, once the measureability of the heavy neutrino
pair-production rate has been established, is the signal
profile.  As explained earlier, Majorana $NN$ production has the
very distinctive signal of a pair of like-sign dileptons and
jets but no
\etsl. In contrast, Dirac $N\overline N$
production is best studied through two pairs (one hard, one
softer) of unlike-sign dileptons and nonzero
\etsl.
The $E_T$- and the rapidity-distributions of the charged leptons
obviously follow two patterns: one for the lepton directly
produced at the $N$-decay  (primary) vertex and the other from
$W$-decay.  While the
former distribution depends on the mass difference $M_N - M_W$,
the latter is much less dependent on $M_N$ and peaks around
$E_T \sim 60$ GeV.

The double $Z$  background can obviously be bothersome in the light of
significant uncertainties possible in $E_T$-measurements at
supercolliders.  But, an acceptance cut of $E_T \geq 30$
GeV would eliminate most of it without affecting the signal
much.  However, if $m_H < 150$ GeV, this last statement does not
hold and one can lose a large part of the signal in imposing an
$E_T$-cut.  A better method of reducing the background might be
to reconstruct the $Z$-events.  A study of the invariant mass
distributions of a pair of oppositely charged final state
leptons in the signal shows that these are {\it not}
peaked around $m_{\ell^+\ell^-} \sim M_Z$.  Thus the imposition
of an acceptance cut of $|m_{\ell^+\ell^-} - M_Z| \gsim
5\Gamma_Z$ would reduce the signal strength only by about
$15~-~20\%$ for $m_N \sim 100~-200~GeV$. On the other hand, this
disposes of almost the entire
$ZZ \longrightarrow \ell^+\ell^-\ell^+\ell^-$ background.

Once the heavy neutrino has been detected, the most obvious
parameter to be determined next is its mass. This can be done
most effectively by looking at the $m_{\ell^+\ell^-}$ distribution
in each hemisphere. For a sufficiently massive $N$
decaying into three massless leptons through charged current
interactions, the
$W$--resonance would  dominate by far. For such a case, the above
invariant mass distribution would have a very sharp cut-off at
$m_{\ell^+\ell^-} = (m_N^2 - m_W^2)^{1/2}$, thus determining $m_N$.

Having established the feasibility of searching for such heavy
neutrinos, let us now turn to the proposal of distinguishing
between Dirac and Majorana ones, first mooted in Ref.
\cite{dicus}.  For the large values of $M_N$, considered here,
the measurement of final state lepton charge signs may not be
easy since these leptons are going to be very energetic.
Nevertheless, the angular correlation between two such leptons
should be measurable.  The authors of Ref. \cite{dicus} employed
the formalism of Tsai  \cite{tsai} to show that the dilepton
angular correlation for the process $gg \arrowz N\overline N, N
\rightarrow \ell X, \overline N \rightarrow \overline{\ell}'X'$ ($X,X'$
unobserved) discriminates between Dirac and Majorana $N$'s.
Such a procedure works in the present case too.  This is
straight forward to see once one realizes that the
discrimination in Ref. \cite{dicus} in fact originates from the
contraction of the $ggZ^*$ vertex and the $Z^*$-propagator to
the $ZN\overline N$ coupling which effectively renders the
latter into a pseudoscalar one.$^{2}$  It follows that, for a
pseudoscalar exchange, the correlation is identical to that of
Ref. \cite{dicus}.  For the scalar-mediated case the structure
of the vertex is a little different, but nonetheless the
discrimination works.

In order to elaborate on the last statement, we go to the
$gg~CM$ frame and write the momenta and spin pseudovectors of
$N$ and $\overline N$ (with velocities $\pm p$) respectively as
\cite{dicus} $$ p_{1,2} = (E,0,0,\pm\beta E),~~S_{1,2} =
\left[\pm\beta (1-\beta^2)^{-1/2}\xi_{1,2~z}, \xi_{1,2~x},
\xi_{1,2~y}, (1 - \beta^2)^{-1/2} \xi_{1,2~z}\right] $$ where
$\vec \xi_{1,2}$ are the spin-vectors in their corresponding
rest frames.  The relations
\be
\barr{rcl}
(1 + S_1\cdot S_2)~(p_1\cdot p_2 + M^2_N) - p_1 \cdot S_2p_2 \cdot S_1
 & = & 2 E^2(1 - \vec\xi_1 \cdot \vec \xi_2),
    \\
(1 - S_1 \cdot S_2)~(p_1 \cdot p_2 - M^2_N) + p_1 \cdot S_2p_2 \cdot S_1
 & = & 2E^2\beta^2(1+\vec \xi_1\cdot \vec\xi_2),
\earr
\ee
follow.  The use of these relations enables us to rewrite
(\ref{H-diff.c.s.}) and (\ref{A-diff.c.s.})
\be
\barr{rcl}
\displaystyle \frac{ d\hat\sigma^D (H)}{d \Omega_N}
   & \equiv & f_H (q^2)~(1+ \vec\xi_1\cdot\vec \xi_2),
   \\
\displaystyle \frac{d\hat\sigma^D(A) }{ d\Omega_N}
   & \equiv & f_A(q^2)~(1 - \vec\xi_1\cdot\vec \xi_2),
\earr
                 \label{dsig in spin}
\ee
where we have lumped all the $\xi$-independent parts into the
factored function $f_{H,A} (q^2)$.

We can further designate the polarization vectors$^{3}$ of $N$
and $\overline N$ as $\vec\omega_1$ and $\vec\omega_2$
respectively and denote the momenta of the final state decay
leptons $\ell(\overline\ell)$, for $N
\rightarrow \ell W, \overline N \rightarrow \overline\ell\overline W$) by
$\vec q_1(\vec q_2)$.  Then, in the notation of Ref.
\cite{dicus}, we can write
\be
\barr{rcl}
\displaystyle \frac{d\Gamma(N \rightarrow \ell X) }{ d\Omega_\ell}
   & = & C_N (\alpha_N - \beta_N\vec q_1\cdot\vec\omega_1),
     \\
\displaystyle \frac{d\Gamma(\overline N \rightarrow \overline \ell\overline X)}
     {d\Omega_\ell}
   & = & C_N (\alpha_N + \beta_N \vec q_2\cdot \vec\omega_2),
\earr
          \label{decay distrib}
\ee
with
\be
\barr{rcl}
C_N &= & \displaystyle
  \frac{\alpha_{EM} B}{ 128 \sin^2\theta_W }~
  \left(1 - \frac{M^2_W}{ M_N^2} \right),
     \\
\alpha_N &=& \displaystyle
   M_N \left( 1 + \frac{M^2_N}{ M^2_W} - \frac{2M^2_W}{M^2_N}
       \right)
       \\
\beta_N & = & \displaystyle
   4 \left( 1 - \frac{~M^2_N}{2 M^2_W} \right),
\earr
               \label{decay param}
\ee
$B$ being the branching ratio for the $W$-decay that is being triggered on
(\eg in the Dirac case it would be the leptonic branching fraction).

Again, as in \cite{dicus}, the use of Eq. (4.26) of Ref.
\cite{tsai} leads using eqs. (\ref{dsig in spin}) and
(\ref{decay distrib}) to the triple
differential cross sections for a Dirac $N$
\be
\barr{rcl}
\displaystyle \frac{d^3\hat\sigma^D(H) }
     { d\Omega_Nd\Omega_\ell d\Omega_{\overline\ell}}
     & = & \displaystyle
      \frac{ f_H(q^2) }{ [\Gamma^D_{TOT}]^2} ~C^2_N
        \left( \alpha^2_N +
               \beta^2_N\vec q_1 \cdot \vec q_2
        \right)
        \and
         \\[2ex]
\displaystyle \frac{d^3\hat\sigma^D(A) }
     { d\Omega_Nd\Omega_\ell d\Omega_{\overline\ell}}
     & = & \displaystyle
      \frac{ f_A(q^2) }{ [\Gamma^D_{TOT}]^2} ~C^2_N
        \left( \alpha^2_N -
               \beta^2_N\vec q_1 \cdot \vec q_2
        \right)
\earr
                   \label{Dirac correl}
\ee
with $\Gamma^D_{TOT}$ as the total decay-width of the Dirac $N$.
In the Majorana case if the lepton charge signs are unobserved,
one must take the sum
\be
 \frac{ d^3\hat\sigma^M }
      { d\Omega_N d\Omega_{\ell^-} d\Omega_{\ell^-} }
+\frac{d^3\hat\sigma^M }
      {d\Omega_N d\Omega_{\ell^+} d\Omega_{\ell^+} }
+\frac{d^3\hat\sigma^M}
      {d\Omega_N d\Omega_{\ell^+} d\Omega_{\ell^-} }.
\ee
 This can
be calculated following Ref. \cite{dicus} to be
\be
  \frac{d^3\hat\sigma^M(H/A) }
       {d\Omega_N d\Omega_\ell d\Omega_{\ell'} }
    =
  \frac{ 16 f_{H,A} (q^2) }{ (\Gamma^M_{TOT})^2}~
     C^2_N\alpha^2_N.
               \label{Majorana correl}
\ee
 In (\ref{Majorana correl}) $\Gamma^M_{TOT}$ is the total
decay width$^{4}$ of the Majorana $N$.

The primary charged dileptons emanating from a pair of decaying heavy
Dirac neutrinos do show a distinctive angular correlation
depending on whether the process involves scalar or pseudoscalar
exchange. For the latter this correlation is identical to that in the
$Z$-exchange case.  Exactly, as in that case, here too the
charged leptons do not show any angular correlation when $N$ is
a Majorana particle.  In reality, all three exchanges ($Z,H$ and
A) will contribute to the process $gg \rightarrow N\overline N$.
But these different contributions correspond to different
initial state angular momenta and parity, \ie different partial
waves, and hence add incoherently.  Thus the basic dilepton
angular correlation discrimination --- yes for Dirac, no for
Majorana --- remains intact.  There is the additional novel
result in this work that the angular correlations for the
Dirac case {\it are} different [{\it vide} eqns. (\ref{Dirac
correl})] for the scalar and pseudoscalar exchange
contributions.  To be sure, these correlations have been
discussed in the $gg~C.M$ frame.  However, given the gluon
density distributions, the transformation of these results to
the $pp~C.M.$ frame in the laboratory can be done by
straightforward Monte-Carlo methods.

Though the above argument holds for $N ( \overline N)
\rightarrow \ell (\overline \ell) + X $, for all $X$, in
practice, for hadronic states $X$, the background is very large
in a $pp$ collider. Thus one has to confine oneself to purely
leptonic $X$. But this could lead to a possible contamination in
the correlation as the argument works only for the charged
leptons coming directly from $N (\overline N)$. One could of
course get around this problem if the two leptons coming from
the same heavy neutrino could be distinguished. Once the mass of
the very heavy neutrino has been determined, $p_T$--distributions
coupled with kinematic considerations then do this job for us.
Of course since we are talking of three body decays, one can
never completely isolate the secondary lepton, but for
sufficiently heavy $N$, there is very little overlap between the
$p_T$ distributions.

In summary, we have presented a feasibility study for finding
very heavy neutrinos at $pp$ supercolliders through direct
pair-production from gluon fusion.  For the mass-range that is
kinematically interesting at such machines, this process
dominates.  We have compared the contributions due to different
propagating bosons in the intermediate state arriving at the
following conclusion.  For a large area of the parameter space,
Higgs-exchange overwhelms the $Z$-contribution, with the
pseudoscalar (if existing) taking the lead.  This is to the
extent that measurable supercollider rates are possible (because
of the heavy top) with only three quark generations --- a fact
that is not true for the $Z$ exchange contribution.   We have
also shown how final state dilepton angular correlation can
discriminate a Majorana heavy neutrino from a Dirac one and, for
the latter, a scalar Higgs exchange contribution from a
pseudoscalar or $Z$-mediated one.  Much of what we say also goes
through for the production of neutral fermions in a
supersymmetric theory.  These and other related issues will be
elaborated elsewhere.

We thank A. Datta, M. Drees and D.P. Roy for helpful
discussions.

\newpage

\centerline{\bf Footnotes}
\begin{enumerate}
\item The function $f(a)=2$ $\left(\sin^{-1} {\sqrt
a/2}\right)^2$ for $0 \leq a \leq 4$ and $2\left(\cosh^{-1}
{\sqrt a/2}\right)^2 -
\pi^2/2 + 2i\pi \cosh^{-1} {\sqrt a}/2$ for $a \geq 4$.

\item In effect, $(q_1 + q_2)^\rho [-g_{\rho\lambda} + (q_1
+ q_2)_\rho (q_1 + q_2)_\lambda M^{-2}_Z]~ \overline
u_N(p_1)\gamma_\mu (g_v + g_A\gamma_5)~v_N(p_2) = 2M_Ng_A
M_Z^{-2} (\hat S - M^2_Z)~\overline u_N(p_1)\gamma_5v_N(p_2)$.

\item $w_{1i} (w_{2i})$ is \cite{tsai} the ratio of the
difference between the number of $N$'s ($\overline N$'s)
polarized along the positive ith axis and that of $N$'s
($\overline N$'s) polarized along the negative ith axis to the
sum of these two numbers.

\item If the field $N$ couples only to the lepton
field $\ell$ and the charged-weak-boson field $W_\mu$, then
$\Gamma^M_{TOT} = 2\Gamma^D_{TOT}$.

\end{enumerate}
\newpage


\newpage

\centerline{\bf Figure Captions}
\begin{enumerate}
\item
   $\sigma( p p \rightarrow N \bar N)$ mediated by a scalar
Higgs and without mixing suppression. \\
   a) Only three generations of quarks with $m_t = 160\;GeV$ \\
   b) Including an additional fourth generation with
      $m_{t'} = m_{b'} = 400\;GeV$.

   The continuous, dotted, dashed and the dot--dashed lines
correspond to $m_H = 100,~200,~300,~{\rm and}~600\;GeV$
respectively.

\item
    Same as Figs. 1a,b, but with  a pseudoscalar Higgs exchange instead.

\item
    Same as Fig. 1a but with $m_H$ as  abscissa.

       The continuous, dotted and the dashed lines
correspond to $m_N = 100,~200,~{\rm and}~300\;GeV$
respectively.

\end{enumerate}

\end{document}